\documentclass[fleqn,10pt]{wlscirep}
\usepackage[utf8]{inputenc}
\usepackage[T1]{fontenc}
\title{Large-scale biometry with interpretable neural network regression on UK Biobank body MRI}

\author[1,*]{Taro Langner}
\author[1,2]{Robin Strand}
\author[1,3]{H\r{a}kan Ahlstr\"{o}m}
\author[1,3]{Joel Kullberg}
\affil[1]{Department of Surgical Sciences, Uppsala University, 751 85 Uppsala, Sweden}
\affil[2]{Department of Information Technology, Uppsala University, 751 85 Uppsala, Sweden}
\affil[3]{Antaros Medical AB, BioVenture Hub, 431 53 Mölndal, Sweden}

\affil[*]{E-Mail: taro.langner@surgsci.uu.se}

\begin{abstract}
In a large-scale medical examination, the UK Biobank study has successfully imaged more than 32,000 volunteer participants with magnetic resonance imaging (MRI). Each scan is linked to extensive metadata, providing a comprehensive medical survey of imaged anatomy and related health states. Despite its potential for research, this vast amount of data presents a challenge to established methods of evaluation, which often rely on manual input. To date, the range of reference values for cardiovascular and metabolic risk factors is therefore incomplete.
In this work, neural networks were trained for image-based regression to infer various biological metrics from the neck-to-knee body MRI automatically. The approach requires no manual intervention or direct access to reference segmentations for training. The examined fields span 64 variables derived from anthropometric measurements, dual-energy X-ray absorptiometry (DXA), atlas-based segmentations, and dedicated liver scans. With the ResNet50, the standardized framework achieves a close fit to the target values (median R$^2 > 0.97$) in cross-validation. Interpretation of aggregated saliency maps suggests that the network correctly targets specific body regions and limbs, and learned to emulate different modalities. On several body composition metrics, the quality of the predictions is within the range of variability observed between established gold standard techniques.

\end{abstract}
\begin{document}

\flushbottom
\maketitle
%
%
\thispagestyle{empty}


\section{Introduction}

As part of the UK Biobank study \cite{sudlow_uk_2015} 100,000 volunteer participants are to be  examined with magnetic resonance imaging (MRI). Among the scheduled imaging protocols is neck-to-knee body MRI, resulting in volumetric images with separate water and fat signal. These scans contain comprehensive information about the anatomy of each subject and are accompanied by a wide range of other collected metadata, spanning anthropometric measurements, questionnaires, biological samples, health outcomes, and more. Many of these properties also express themselves in the morphology of the human body and could potentially be inferred with machine learning. Techniques involving neural networks for image-based regression have been previously proposed for the analysis of brain MRI for detection of premature ageing \cite{cole_predicting_2017-1}, early symptoms of Alzheimers disease \cite{ding2018deep} and mental disorders \cite{shahab2019brain}. In heart MRI, related approaches were able to perform measurements of volumes and wall thicknesses of the heart \cite{xue2017direct}. Similarly, analyses of retinal fundus photographs showed that neural networks were able to leverage image features for the prediction of properties including age, gender, smoking status and blood pressure \cite{poplin2018prediction}. Many of these findings were unexpected as the underlying features are often not easily accessible even to human experts. \\
Research in metabolic and cardiovascular disease has led to increased interest in strategies for the automated analysis of body composition \cite{thomas2013whole}. Individualized measurements of fat and muscle compartments in the body have the potential to provide new insight into the development of various medical conditions at greater detail than analyses based on anthropometric measures such as the body mass index (BMI) \cite{prentice2001beyond}. The amount of visceral adipose tissue in particular varies substantially between individuals and is directly related to cardiac and metabolic risk \cite{neeland2013associations}. A more fine-grained analysis is of interest in research such as within the UK Biobank study itself \cite{linge2018body} but also as a potential tool for disease screening and individualized treatments. Several imaging techniques exist for the measurement of body fat, including computed tomography (CT) and dual-energy X-ray absorptiometry (DXA) \cite{kaul2012dual} based on two-dimensional coronal projections. Chemical-shift encoded water-fat MRI acquires separate volumetric water and fat signal images which have the potential to allow for measurements without ionizing radiation, but can be challenging to evaluate. Various methods have been proposed for the delineation of individual adipose tissue depots in these images\cite{hu2016segmentation}. Among other techniques, automated image analysis with convolutional neural networks for segmentation has become an established technique for images of this kind \cite{doi:10.1002/mrm.28022, langner2019fully} as well as for CT images \cite{weston2018automated, wang2019effective}. However, these systems learn to perform segmentation from training data in the form of reference segmentations, which must accordingly be carefully prepared, often with substantial amounts of manual guidance. \\

In this work, automated biometry is performed by training neural networks for image-based regression on UK Biobank neck-to-knee body MRI. The proposed approach extends a previously presented method for age estimation \cite{Langner2019} and requires no manual intervention or direct access to ground truth segmentation images. Instead, arbitrary numerical values can be inferred, ranging from anthropometric measurements to body composition metrics from dual-energy X-ray absorptiometry (DXA), multi-atlas-based MRI segmentations, dedicated liver scans and various other sources. The goal of this approach is to approximate all of these measurements with a fast and accurate, fully automated technique from the MRI data. \\

\noindent The following contributions are made:
\begin{itemize}
	\item Extension of a framework for age estimation from UK Biobank neck-to-knee body MRI \cite{Langner2019}
	\item Inference of 64 biological metrics (beyond just age)
	\item Design of an optimized and standardized configuration
	\item Extensive validation of both framework and predictions
	\item Aggregated saliency analysis \cite{Langner2019}
	
\end{itemize}

To our knowledge, no comparable technique with convolutional neural network regression has been previously applied to neck-to-knee or whole-body MRI for inference of biological metrics other than age. Essential code, documentation and supplementary material has been made available for reproducibility and further use. \cite{github_biometry}

\section{Methods}

A fixed configuration of a convolutional neural network for image-based regression was trained in cross-validation on two-dimensional representations of the neck-to-knee body MRI. For each of the 64 examined properties, the network was evaluated based on the generated predictions and saliency maps which highlight relevant image features.

\subsection{Image data}

\begin{figure}

\begin{minipage}[t]{.45\textwidth}	
	\includegraphics[width=\textwidth]{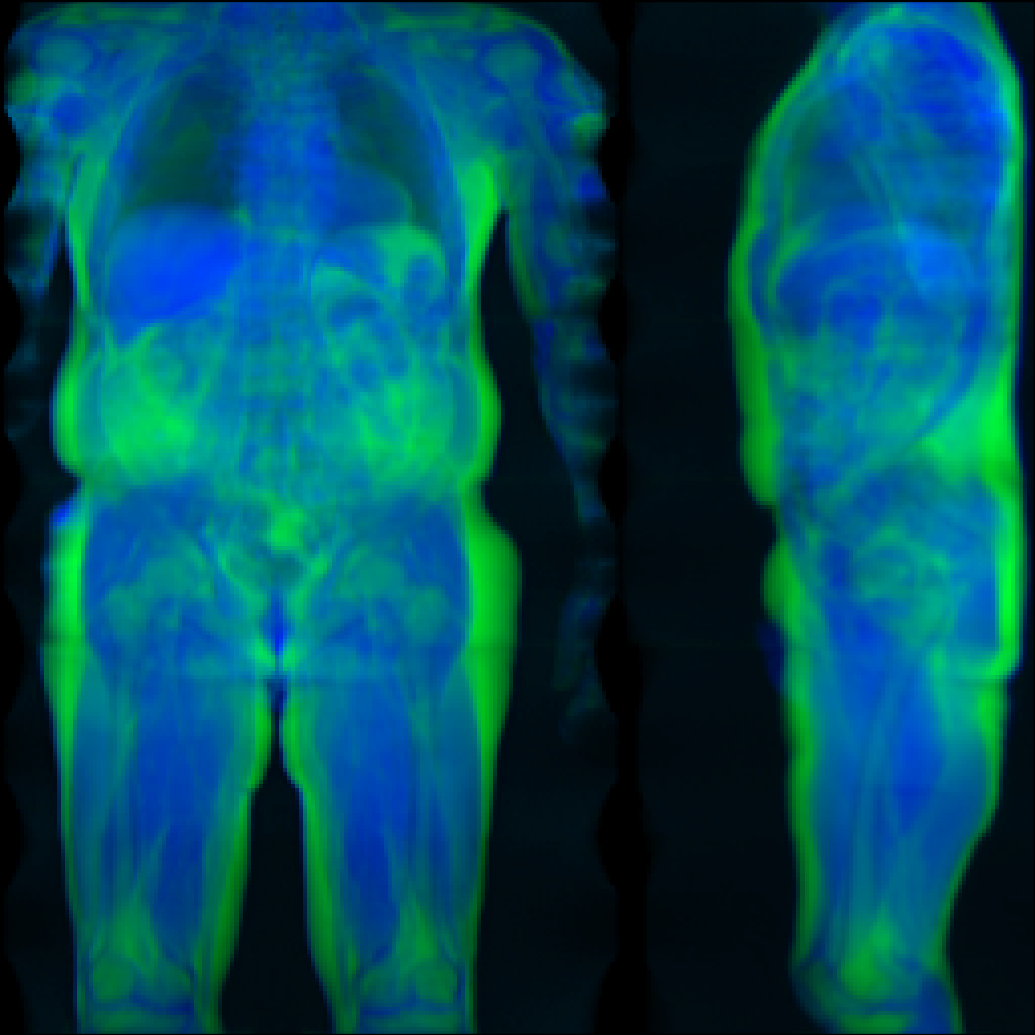}
	\caption{Two-dimensional representation of the volumetric MRI data, which serves as input to the neural network. The water (blue) and fat (green) signal images are projected along the coronal and sagittal plane and combined as color channels.}
	\label{fig_input}
\end{minipage}
\begin{minipage}[t]{.45\textwidth}	
	\includegraphics[width=\textwidth]{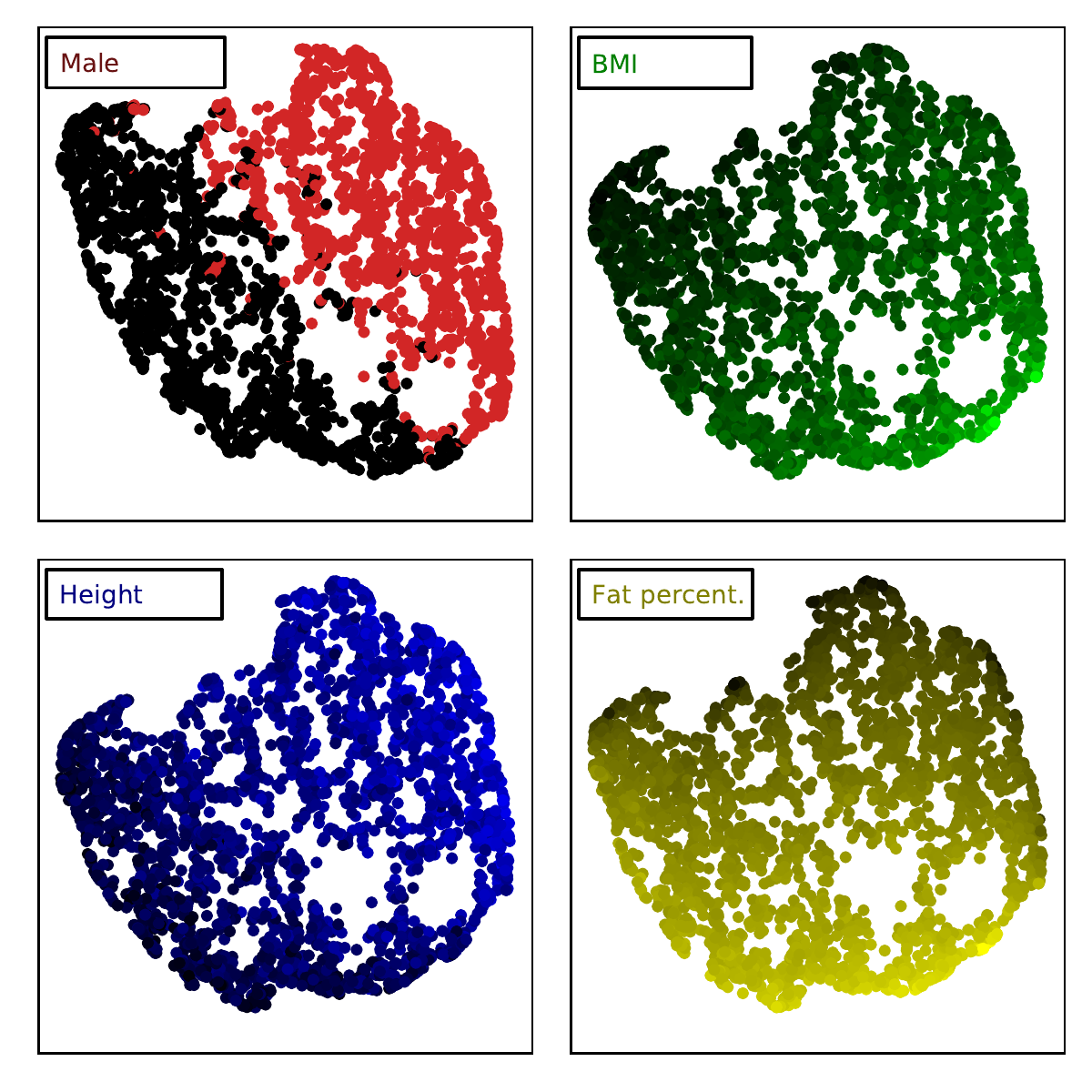}
	\caption{Visualized feature space of body composition, with 64 chosen fields of 3,000 subjects. The 2d UMAP projection \cite{mcinnes2018umap} minimizes the distance between similar subjects, each represented by one dot.	For each subplot, the named variable was removed before dimensionality reduction and used for colouring, with brighter intensity for higher values. The sexes (field 31, 0 for females and 1 for males) distinctly divide in two hemispheres whereas the ranges of BMI (field 21001), height (field 12144) and body fat percentage (field 23281) vary systematically.}
	\label{fig_umap}

\end{minipage}

\end{figure}

Of the 100,000 MRI scans planned by the UK Biobank study, 32,323 were made available for the experiments in this work as part of application 14237. UK Biobank recruitment was organized by letter from the National Health Service and the vast majority of participants (94\%) self-reported white British ethnicity in the initial assessment visit. All scans were acquired by the UK Biobank at three different centres in the United Kingdom in an imaging time of about six minutes each, using a dual-echo Dixon technique \cite{hu2013quantitative} on a Siemens Aera 1.5T device. The resulting image data typically covers the body from neck to knee in six separate stations, whereas the arms and other parts of the body that extend laterally are usually not visible or subject to heavy distortion and artefacts \cite{west_feasibility_2016}. For the experiments in this work, those scans that contained water-fat swaps and other artefacts such as excessive noise, unusual positioning and artificial knee replacements were excluded by visual inspection of the projections by one operator, leaving 31,172 images for training and validation. The volumetric scan stations for a given subject were resampled to a resolution of 2.23 mm $\times$ 2.23mm $\times$ 3mm and fused into a volume of 370 $\times$ 224 $\times$ 174 voxels. This MRI volume was then cropped and compressed into a two-dimensional format of slightly lower resolution, showing a frontal and lateral projection of mean intensity, with a separate image channels for the water and fat signal. In this format, each subject was accordingly represented by a two-channel image of 256 $\times$ 256 pixels, as seen in Fig. \ref{fig_input}, stored in 8bit format for easier processing by the neural network.

\subsection{Biological metrics}

From the thousands of non-imaging properties collected in the UK Biobank study, a subset of 64 fields with relevance for cardiovascular and metabolic disease was chosen. More than half of the chosen fields are measurements of body composition by DXA imaging\cite{harvey2013osteoporosis, kaul2012dual}, comprising mass and percentages of fat and lean tissue in the abdomen, trunk, arms and legs. The second largest group of measurements is based on multi-atlas segmentations of the neck-to-knee body MRI itself\cite{borga2015validation,west_feasibility_2016,borga2018advanced} and describe volumes of adipose tissue depots and muscle groups in the abdomen, trunk and thighs. An additional group of fields contains the basic features of age, sex (1 for male, 0 for female), height, and weight. Due to privacy concerns, the age could only be calculated to an accuracy of about 15 days, based on the year (field 34) and month of birth (field 52) as well as the MRI scanning metadata (field 20201)\cite{Langner2019}. 
The last group of fields contains values such as circumferences of the hip and waist, BMI, the percentage of fat accumulated in the liver, determined by dedicated liver MRI \cite{wilman2017characterisation}, the pulse rate on the imaging visit, and the measured grip strength of the right hand, which is often used as an biomarker for cardiovascular health. Of the 32,323 imaged subjects, only 3,048 have valid entries for all of the chosen fields. These subjects serve as a basis for the saliency analysis, described later in this chapter. A feature space of the 64 chosen metadata fields for these subjects is also visualized in Fig. \ref{fig_umap} and showcases some of the underlying patterns relating to sex and body composition.

Using one standardized configuration, a dedicated neural network was trained to predict each of these 64 measurements separately. Each of them was evaluated in 7-fold cross-validation, so that all of those subjects with a valid entry for the given measurement were split into 7 subsets of equal size. By exempting each subset in turn from training and using it to make predictions which could then be compared to the reference, the network was effectively validated against all subjects without being able to memorize their values in training.

\subsection{Network configuration}

For each of the chosen fields a separate convolutional neural network was trained for regression in 7-fold cross-validation. The entire configuration of the network was fixed and no attempt was made to achieve better performance by tuning the network architecture or other parameters. Each unique training sample represents one subject and consists of two-dimensional format as extracted from the MRI data as input image and their field entry in the UK Biobank as numerical ground truth target value.

The neural network is a computational model that uses millions of variable parameter weights to convert an input image into one or more numerical output values. During training, it can learn to perform a certain task by making image-based predictions for samples with known reference values. The difference between prediction and reference is quantified by a loss function, and mathematical optimization involving its gradient adjusts the network parameters. In this way, parameter values can be learned that define convolutional image filters for extraction of relevant gradients, corners and edges from the image, which are subsequently formatted into increasingly abstract features that enable the network to infer the desired measurement. This process is entirely data driven and fully automated.

The previously presented regression pipeline \cite{Langner2019} for age estimation was optimized in several ways in order to process all of the chosen fields in a viable time frame. The main change consists in replacing the VGG16 architecture \cite{simonyan_very_2014} with the more lightweight ResNet50 \cite{he_deep_2016}. Furthermore, all numerical target values were standardized by subtracting the mean value and dividing by the standard deviation, as the ResNet50 proved more sensitive to variation in target scaling and shifts. This step resulted in faster convergence and improved stability, so that the total number of iterations could be vastly reduced from 80,000 iterations to just 6,000. To alleviate a tendency of the network to overfit in the final 1,000 iterations, the learning rate of $0.0001$ in this phase is reduced by factor ten, typically resulting in a further slight increase in accuracy. Compared to the original configuration, the total training time for a given field was thus reduced by about factor 30, while reaching comparable accuracy. The original batch size of 32 and augmentation by random translations of up to 16 pixels were retained, with the nearest pixel values being repeated at the borders. All networks were trained on a Nvidia GTX 1080 Ti 11GB graphics card in the framework PyTorch with a mean squared error loss, the optimizer Adam, and parameters pretrained on ImageNet. Each split required less than 25 minutes of training time.

These design choices were made based on preliminary results for three representative fields: Age, liver fat (field 22402) and visceral adipose tissue volume (VAT) (field 22407). All presented results were achieved with this exact network configuration, without early stopping, hyperparameter tuning, or any other attempt to adapt to individual fields for better performance.

\subsection{Evaluation}

The chosen fields range from volumes to circumferences and simple binary labels, all treated as continuous numerical values. The neural network was trained to predict these values in regression, thereby emulating the reference, and the coefficient of determination R$^2$ is reported to rate the quality of fit, ranging from $1.0$ for a perfect fit to negative values where the non-linear network model performs worse than simply estimating the mean. Additionally, the 95\% limits of agreement (LoA) and the mean absolute error (MAE) are provided.
In some cases the network output was thresholded to mimic a classification, with a threshold of $0.5$ for prediction of sex and $5.5\%$ for fatty liver disease. Without taking the exclusion criteria into account, the reference liver fat values of 898 of 4219 subjects exceed this threshold. For the prediction, an area under curve (AUC) of the receiver operating characteristic (ROC) curve was calculated.

In some cases, competing measurements of the same property are available from several reference methods, so that their mutual agreement can be compared to the network performance. In the scope of this work, only the atlas-based MRI segmentations \cite{borga2018advanced} and measurements from DXA \cite{harvey2013osteoporosis} are considered in this regard. Both methods examine different regions of interest and therefore show systematic differences. The MRI-based values were therefore first fit to the DXA values by linear regression before reporting their agreement in this analysis. Similarly, many fields describe features specific to the left and right side of the body. Again, the network performance can be put into the context of this inherent bilateral symmetry, but this analysis is abbreviated to report Pearson's coefficient of correlation $r$ only.

In addition to statistical measures, an interpretation of the criteria learned by the network can be attempted with saliency analysis. For each input image, a heat map of relevant image features can be generated using guided gradient-weighted class activation maps \cite{selvaraju_grad-cam:_2017, uozbulak_pytorch_vis_2019}. The resulting visualizations were combined by co-registration of subjects \cite{ekstrom_fast_2018}, yielding aggregated saliency maps that describe which image regions on average had the highest impact on the network prediction\cite{Langner2019} for an entire cohort of subjects. Each saliency map was generated by the one network that used the corresponding subject as a validation sample in cross-validation. When visualized, the saliency intensities were squared and overlaid as a heat map over the water signal image, without any further post-processing or manual adjustment.

Some properties could be trivial to predict due to strong correlations with simple non-image features such as age and weight. We therefore also provide the results of multiple linear regression based on the age, sex, height and weight as a baseline for comparison with the neural network performance.

\section{Results}

\begin{figure}
	\centering
	\includegraphics[width=\textwidth]{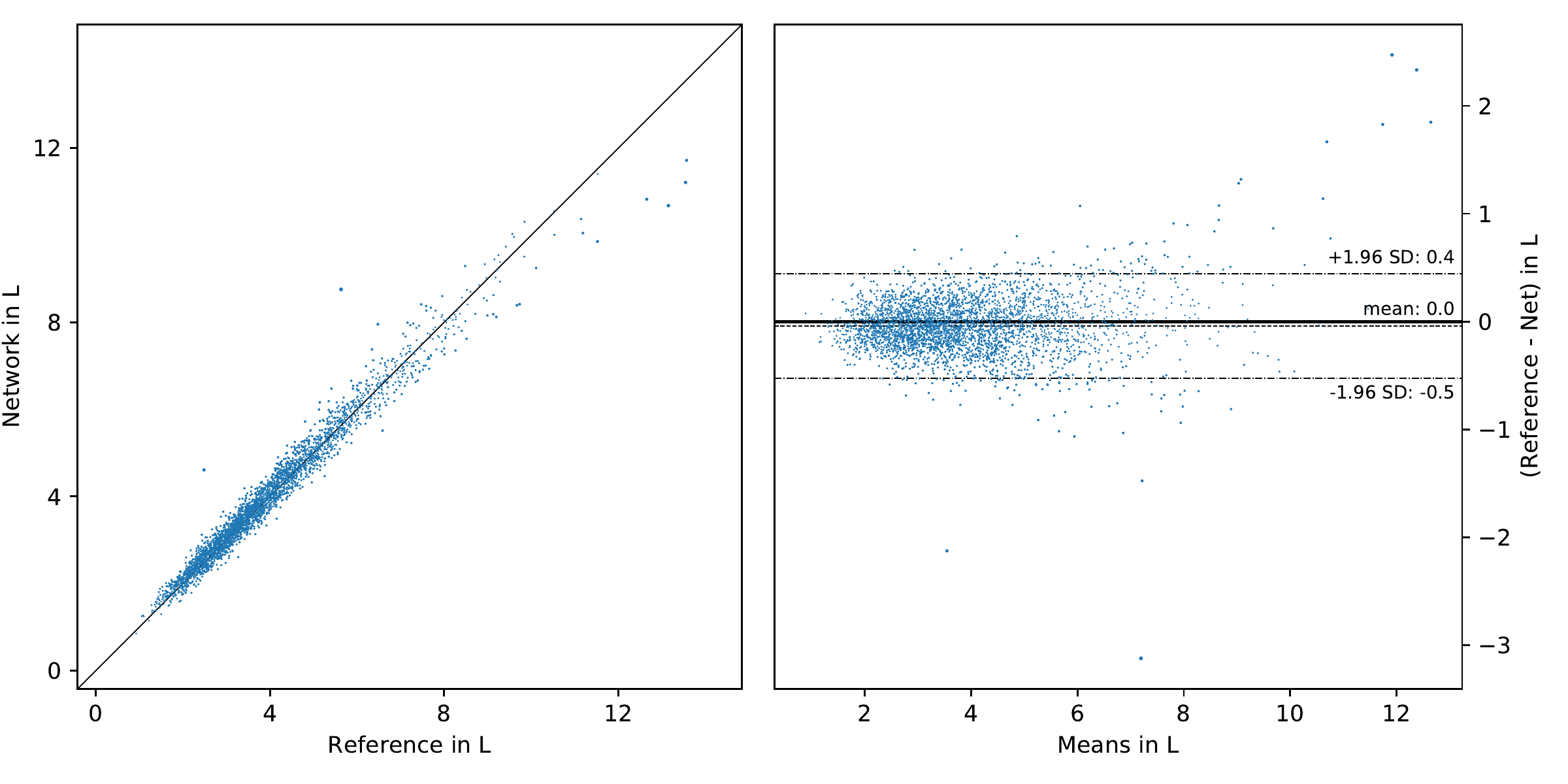}	
	\caption{Visualization of the median result with R$^2=0.972$, the inference of fat mass in the left leg as measured by dual-energy x-ray absorptiometry (DXA) (field 23266). The diagonal line represents a hypothetical perfect fit, whereas dashed lines represent the 95\% limits of agreement (LoA). Point sizes vary for better visibility.}
	\label{fig_corr}
	
\end{figure}

A close regression fit is achieved on almost all examined fields. On average, less than 3\% of variability in the reference measurements remains unexplained by the network output alone (median R$^2=0.972$) and the linear regression baseline was outperformed in all cases. The field with median fit is shown in Fig. \ref{fig_corr}, and more plots for all fields are available in the Supplementary Material\cite{github_biometry}. Table \ref{tab_base} lists the basic fields with a MAE of about 2.5 years for age, 0.8kg for body weight and 1.7cm for height. When thresholded, the classification accuracy for the prediction of sex reaches $99.97\%$, so that only 10 of 31,172 subjects were misclassified. 

Some of the most accurate predictions were made for body composition as measured by atlas-based segmentation on MRI (median R$^2=0.987$), with a corresponding MAE of 140 mL for visceral adipose tissue (VAT), 220 mL for subcutaneous abdominal adipose (ASAT), and 180 mL for total thigh muscle volume. Additional statistical metrics for these fields and others including those from DXA and liver fat are provided in Supplementary Tables \ref{tab_main}, \ref{tab_dxa_trunk}, \ref{tab_dxa_arm}, and \ref{tab_dxa_leg}. The lowest performance was achieved on grip strength and pulse rate, where the network nonetheless managed to make a weak, image-based prediction from the MRI. When thresholded at $5.5\%$ to identify subjects with high liver fat, the predictions reached an accuracy of $90\%$, with a sensitivity of $73\%$, specificity of $95\%$ and an AUC-ROC of $0.943$. Even though the arms are usually not visible in the images, the network succeeded in estimating the grip strength of the right hand with an MAE of about 5kg and furthermore gave a rough estimate of the pulse rate. 

\begin{table*}[]
	\begin{center}
		\caption{Inference of basic fields}
		\label{tab_base}
	\begin{tabular}{	
	llrl
	r@{\hskip 0.05cm}c@{\hskip 0.05cm}r
	r@{\hskip 0.05cm}c@{\hskip 0.05cm}l | @{\hskip 0.05cm}
	r c r}
		\hline
		field & name & N & unit &  [min&,&max] & mean &$\pm$& SD & MAE & LoA & R$^2$\\ 
		\hline	
		/ & age & 31172 & years & [44.6 & , & 82.3] & 63.9 & $\pm$ & 7.5 & 2.46 & (-5.85 to 6.31) & 0.829 \\
		31 & sex & 31172 & (male = 1) & [0.0 & , & 1.0] & 0.5 & $\pm$ & 0.5 & 0.00 & (-0.04 to 0.04) & 0.999 \\
		12144 & height & 31172 & cm & [140.0 & , & 204.0] & 170.2 & $\pm$ & 9.4 & 1.70 & (-4.78 to 4.44) & 0.938 \\
		21002 & weight & 30382 & kg & [39.6 & , & 169.2] & 76.5 & $\pm$ & 15.1 & 0.78 & (-1.95 to 2.24) & 0.995 \\	
		
			\hline
			\hline
			\multicolumn{13}{l}{*SD: Standard deviation, MAE: Mean absolute error, LoA: Limits of agreement.} \\		
		\end{tabular}
	\end{center}
\end{table*}

\subsection{Saliency analysis}
\begin{figure}
	\includegraphics[width=\textwidth]{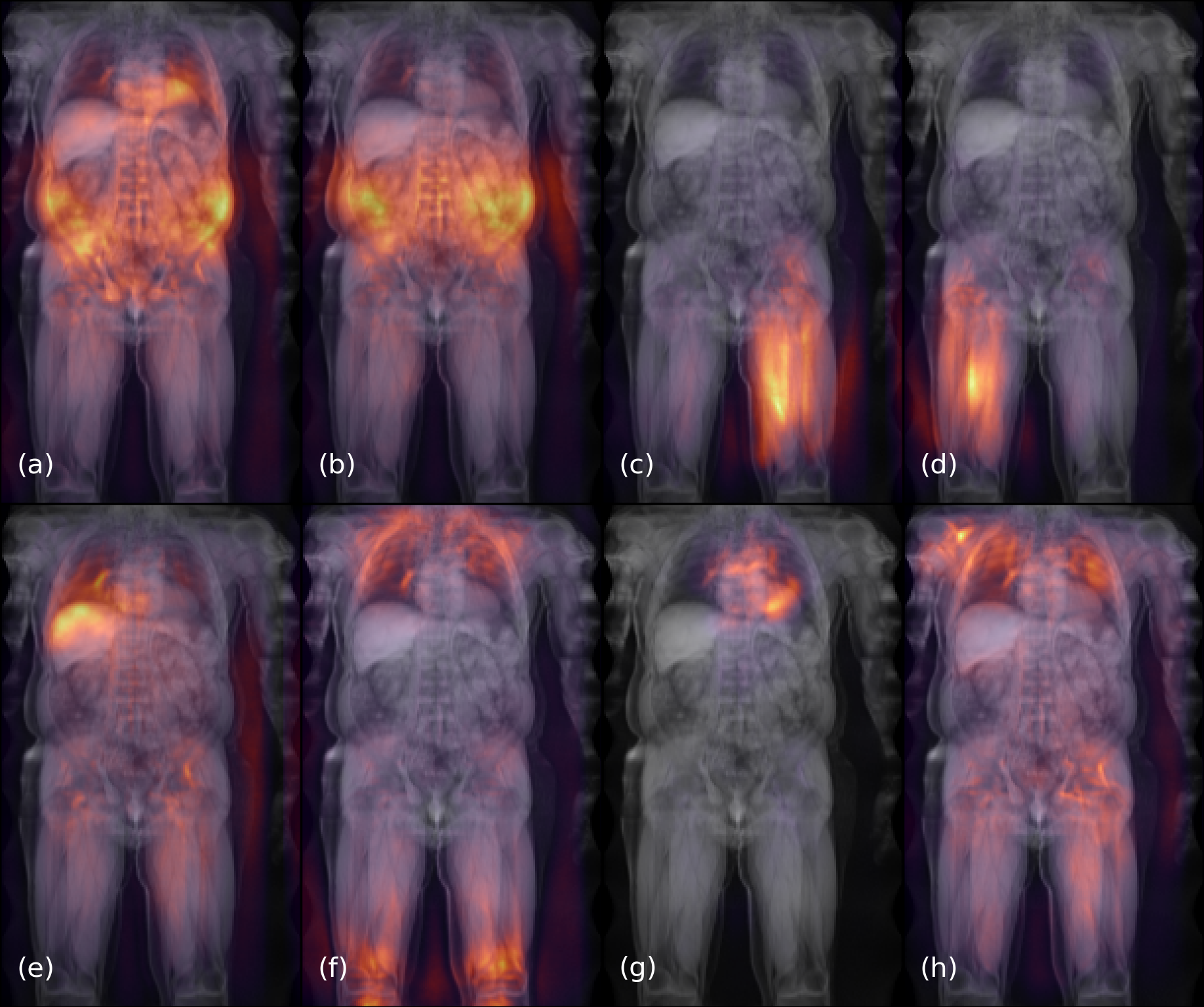}	
	\caption{
		Aggregated saliency of about 3,000 subjects for: VAT as derived from atlas-based MRI segmentations (field 22407) (a)	or DXA (field 23289) (b), muscle volumes of the anterior left (field 22405) (c) and right thigh (field 22403) (d), liver fat (field 22402) (e), BMI (field 21001) (f), pulse rate (field 102) (g) and grip strength (field 47) (h). 
		The network appears to emulate regions of interest used by different modalities and correctly targets specific limbs and organs.}	
	\label{fig_sal}
\end{figure}

Examples for saliency maps generated by the network are shown in Fig. \ref{fig_sal}. The saliency indicates that the network on average correctly targets specific structures on the left or right side of the body. Moreover, the estimate of liver fat appears to be mostly based on image areas with actual liver tissue, whereas the prediction of the pulse rate takes into account features of the heart. The BMI appears to be mostly estimated from the knees and lungs, and the grip strength of the right hand is inferred from features of the corresponding side of the upper body. Complete visualizations of all saliency maps are provided in the Supplementary Material\cite{github_biometry}.

\begin{table*}
	\begin{center}
		\caption{Agreement between reference methods}
		\label{tab_compare}		
		\begin{tabular}{lll|ccc|ccc}
			\hline
			property & field$_{MRI}$ &  field$_{DXA}$ & N & term & unit & MAE & LoA & R$^2$\\ 
			\hline
			VAT & 22407 & 23289 & 4494 & $0.42 x - 0.3$ & L & 0.17 & (-0.45 to 0.45) & 0.939	\\
			TotalTrunkFat &22410 & 23284 & 4538 & $1.35 x + 0.6$ & kg & 0.75 & (-1.93 to 1.93) & 0.971  \\
			TotalFatTissue & 22415  & 23278 & 4326 & $1.23 x -0.2$ & kg & 0.85 & (-2.30 to 2.30) & 0.981 \\
			TotalLeanTissue & 22416  & 23280 & 4326 & $1.88 x + 1.9$ & kg & 1.68 & (-4.46 to 4.46) & 0.937 \\
			\hline
			\hline
			\multicolumn{9}{l}{* Transforming measurements from atlas-based MRI segmentations of N subjects to their corresponding values } \\
			\multicolumn{9}{l}{ from DXA yields the listed linear regression term, mean absolute error (MAE), and limits of agreement (LoA).} \\
			\multicolumn{9}{l}{ In all cases, this agreement is exceeded by the accuracy of their individual network predictions.}
		\end{tabular}
	\end{center}
\end{table*}

\subsection{Agreement between modalities}

Measurements from DXA are compared to those derived from atlas-based segmentations of the MRI in Table \ref{tab_compare}. Each listed comparison yielded lower agreement between these reference methods than achieved by the specific network predictions, evaluated in Supplementary Tables \ref{tab_main} and \ref{tab_dxa_trunk}. Although only a one-way fitting of MRI to DXA is shown, this analysis was performed in both directions and yielded average LoA between both methods that are $70\%$ wider on average than the LoA between each field and its network predictions.

\subsection{Bilateral symmetry}

In some cases the accuracy of the network predictions also exceeds the inherent, bilateral symmetry of the human body. For a given property, one limb is accordingly more dissimilar to the opposite limb than to its prediction by the network. A field-wise comparison with Pearson $r$ is reported in Supplementary Table \ref{tab_symmetry}. For atlas-based measurements from MRI, the average bilateral correlation for the anterior and posterior thigh muscle volume amounts to $r = 0.979$. The network predictions correlate more strongly with the left- and right-specific measurements for an average $r = 0.989$. For DXA, however, the specific prediction accuracy of the network is lower than the bilateral symmetry, with averages of \mbox{$r = 0.975$ vs $r=0.954$} for the arms and \mbox{$0.987$ vs $0.983$} for the legs. 
Although some individuals show strong unilateral atrophy, this effect is not just due to outliers. The fact that the network learned to specifically target either side of the body is also visible in the saliency maps of Fig. \ref{fig_sal} and occurs in both the DXA and MRI-based fields.

\section{Discussion}

The neural network configuration showed robust performance and closely emulated the chosen measurements by image-based regression on the MRI data, with a median R$^2$ above 0.97. It not only learned to accurately estimate volumes and circumferences from the simplified, two-dimensional image format, but also to emulate different modalities and make measurements specific to either side of the body. The linear regression baseline was outperformed in all cases and indicates that most of these properties can not be trivially deduced from the basic characteristics of age, sex, height, and weight. 

When used to infer metrics related to body composition, the network yielded more faithful approximations of the atlas-based measurements from MRI or DXA than obtained by substituting these two reference methods for each other. This was still the case even after fitting both reference methods to each other with linear regression. The agreement for both modalities on the UK Biobank reported in previous work \cite{borga2018advanced} yielded similar error bounds, for a sample with considerable overlap to the subjects examined here. The atlas-based method on MRI has also been previously compared to an alternative method based on T1-weighted images \cite{borga2015validation}, yielding LoA for VAT, ASAT and total trunk fat that are on average more than twice as wide as those relative to the network here. The variability between these two established reference methods can largely be accounted for by differing regions of interest. Whereas the atlas-based method measures VAT up to the thoracic vertebrae Th9 \cite{west_feasibility_2016}, DXA defines VAT as ranging from the top of the iliac crest up to 20\% of the distance to the base of the skull \cite{borga2018advanced}. This is reflected in the saliency maps of Fig. \ref{fig_sal}, indicating that the network correctly learned to emulate the different criteria, based on the numerical target label alone.

Many of the most accurate predictions were made for the atlas-based measurements on MRI, where the accuracy of the network also exceeds the inherent similarity in muscle volumes between the left and right leg. There are several possible explanations for this. In contrast to the DXA-based values, these reference measurements were originally performed on the same MRI data that served as a basis for the presented method. The lack of outliers in the reference suggests high quality, closely representing an objective truth that is contained in these images. Furthermore, all images with ground truth values passed the quality control steps applied by the reference. The network was accordingly trained and evaluated on samples that were preselected regarding suitability for body composition analysis. The measurements of the arms and legs from DXA, in contrast, contain outliers and are often based on anatomy that is not entirely contained in the field of view. Including additional imaging stations that would cover the lower legs and head could lead to both more robust inference and better agreement with the DXA measurements, but was rejected during the original study design due to the prohibitive total increase in scan time\cite{west_feasibility_2016}. Future studies may benefit from a less targeted acquisition and instead choose to collect less restricted, more comprehensive data, as increasingly powerful tools for automated analysis become available.

Despite being able to use the same MRI data and producing similar measurements, the proposed technique and the atlas-based reference method\cite{west_feasibility_2016} differ substantially in their approach. The network generates no segmentations for manual refinement or quality control. It furthermore requires hundreds or thousands of labelled ground truth images for training and would likely require retraining for different imaging devices and demographics. The atlas-based method relies on just 31 prototype subjects and has been credited for robustness towards different imaging devices and field strengths.
In turn, the network can analyse several scans within just seconds instead of minutes and requires no manual intervention or guidance, so that it can easily be scaled to process tens of thousands of subjects. Even though no segmentations are generated, there is also no restriction on using only segmented images as input, but instead arbitrary numerical target labels can be used. This makes it possible to examine more abstract properties, such as grip strength and pulse rate, and to link them to relevant anatomical regions by saliency analysis.

One limitation of this work consists in the lack of an independent test set. This means that it remains unclear whether the already trained networks would reach similar performance on data from other studies and sources. As the used data has been gathered at three different imaging centres, it at least appears that the protocol can be reproduced sufficiently well at different sites to allow for robust performance on future UK Biobank images of the same population. When applied to data from other studies, such as for example the whole-body MRI scans of the German National Cohort \cite{bamberg2015whole}, systematic differences in subject demographics, scanning device or protocol are likely to limit the performance however, and retraining of the networks would almost certainly be necessary. 
The lack of an independent test set might also raise concerns about the network configuration being excessively adapted to the given data. It could be assumed that the repeated runs of the cross-validation during the preliminary experiments may have resulted in design choices that merely represent a coincidental optimum on the cross-validation data itself, with low ability to generalize and possible dependence on confounding factors in the images. However this effect is unlikely to play a significant role since all design choices were based on preliminary experiments on the fields for age, liver fat (field 22402) and VAT (field 22407) only. The resulting configuration is robust without any individual adjustment for a large variety of measurements with tens of thousands of subjects, so that it is exceedingly unlikely that the high performance is coincidental or based on simple confounding effects alone. 

Many properties could potentially be predicted with greater accuracy by using customized image formats and more training samples. The resampled, two-dimensional projection effectively compresses the volumetric MRI data by factor 220 and is furthermore encoded to 8bit only. Despite the computational benefits there is no reason to assume that this format is optimal for all examined fields. Among its limitations, the separately normalized water and fat signal only enable an indirect inference of fat fraction values. When inferring these values for certain tissues and organs, the signals are furthermore conflated along the axis of projection. Future work will explore ways to make this information more accessible to the network, which is likely to benefit especially the inference of liver fat. Despite the limitations of the dual-echo Dixon technique for this purpose \cite{kukuk2015comparison}, these improvements may ultimately yield higher agreement than observed between other methods such as biopsy and magnetic resonance spectroscopy\cite{roldan2010vivo}.

When compared to the previous configuration for age estimation, \cite{Langner2019} the network for age was trained in cross-validation with about $28\%$ more data. The mean absolute error accordingly decreased as expected, from a previous 2.49 years to 2.46 years, roughly following the previously reported relationship between performance and quantity of training data. The ResNet50 performs similar to the VGG16 when using standardization of the target values, but at far higher speed. Its main disadvantage consists in more diffuse saliency maps, possibly due to the final average pooling layer.

The results show that the presented approach can leverage the two-dimensional representation of MRI image data to estimate not only the age but also to emulate a wide range of other measurements for subjects of the UK Biobank. Given only an abstract, numerical target value and the vast amount of images, the regression network learned to identify the correct body region, tissue or limb as used by the reference methods. In its current form the method could be used as a fully automated tool for approximation of missing values for those subjects who have not yet undergone all of the planned examinations. These estimates could then serve for quality control and as a basis for preliminary analyses, months or years before the established gold standard methods have been fully applied. Future work will consist in making the results accessible to the medical community and improving individual measurements with specialized input formats and network configurations, as well as exploring the limits of which other, more abstract properties can be predicted from these scans. Similar approaches could potentially enable the prediction of more variables such as blood biochemistry, disease states, and genetic markers.

\section{Conclusion}

The neural network can perform fully automated inference on the UK Biobank MRI data and learned to emulate measurements from DXA, atlas-based segmentations, dedicated liver scans and more in a fast and lightweight, standardized configuration. Saliency and correlation analysis indicate that the network can specifically target the left and right side of the body and identify relevant organs and body regions. Given enough training data for a given demographic and a standardized imaging protocol, further development may ultimately enable fully automated measurements of a wide range of biological metrics from a single 6-minute neck-to-knee body MR image.

\section*{Acknowledgements}

This work was supported by a research grant from the Swedish Heart- Lung Foundation and the Swedish Research Council (2016-01040, 2019-04756) and used the UK Biobank resource under application no. 14237.

\section*{Author contributions statement}

T.L. wrote the main manuscript text and conducted the experiments, R.S. revised the manuscript, H.A. and J.K. supervised data access from the UK Biobank and contributed to the statistical evaluation. All authors read and approved the final manuscript.
 
\section*{Additional information}

\bibliography{references}

\pagebreak

\section{Supplementary Material}
\renewcommand{\tablename}{Supplementary Table}
\setcounter{table}{0}

\textbf{Title: Large-scale biometry with interpretable neural network regression on UK Biobank body MRI}\\
Authors: Taro Langner, Robin Strand, H\r{a}kan Ahlstr\"{o}m, Joel Kullberg

\vspace{0.5cm}
\noindent
Evaluation details for all remaining targets are given in Supplementary Tables \ref{tab_main}, \ref{tab_dxa_trunk}, \ref{tab_dxa_arm}, and \ref{tab_dxa_leg}, whereas Supplementary Table \ref{tab_symmetry} lists Pearson's correlation coefficients for the similarity between measurements of symmetrical body parts to each other and to their predictions by the network.

\begin{table*}[htp]
	\begin{center}
		\caption{Inference of main fields}
		\label{tab_main}		
		\begin{tabular}{
				ll@{\hskip 0.05cm}rl@{\hskip 0.2cm} 
				r@{\hskip 0.05cm}c@{\hskip 0.05cm}r@{\hskip 0.1cm} 
				r@{\hskip 0.05cm}c@{\hskip 0.05cm}l  | @{\hskip 0.05cm}
				r@{\hskip 0.05cm} c@{\hskip 0.05cm} r|l}
			\hline
			field & name & N & unit &  [min&,&max] & mean &$\pm$& SD & MAE & LoA & R$^2$ & R$^2_{lr}$\\ 
			\hline	
			& \textbf{DXA:} & &&&&&&&&&&& \\
			
			23279 & TotalFatFreeMass & 4544 & kg & [7.5 & , & 84.3] & 49.9 & $\pm$ & 10.2 & 0.79 & (-2.05 to 1.98) & 0.990 & 0.908 \\
			23278 & TotalFatMass & 4544 & kg & [2.7 & , & 76.1] & 26.1 & $\pm$ & 9.0 & 0.56 & (-1.36 to 1.59) & 0.993 & 0.894 \\
			23280 & TotalLeanMass & 4544 & kg & [6.8 & , & 80.3] & 47.3 & $\pm$ & 9.7 & 0.82 & (-2.07 to 2.11) & 0.988 & 0.904 \\
			23281 & TotalTissueFat & 4544 & \% & [11.8 & , & 58.4] & 35.2 & $\pm$ & 8.0 & 0.67 & (-1.67 to 1.68) & 0.988 & 0.740 \\
			23282 & TotalTissueMass & 4544 & kg & [9.5 & , & 154.0] & 73.3 & $\pm$ & 14.5 & 0.79 & (-2.07 to 2.17) & 0.994 & 0.993 \\
			23289 & VatVolume & 4498 & L & [0.0 & , & 6.6] & 1.3 & $\pm$ & 1.0 & 0.12 & (-0.30 to 0.33) & 0.972 & 0.718 \\
			
			&&&&&&&&&&&&& \\
			& \textbf{MRI:} & &&&&&&&&&&& \\
			
			22403 & AnteriorThighMuscleR & 5662 & L & [0.6 & , & 3.7] & 1.7 & $\pm$ & 0.5 & 0.06 & (-0.15 to 0.15) & 0.975 & 0.796 \\
			22404 & PosteriorThighMuscleR & 5662 & L & [1.4 & , & 6.2] & 3.4 & $\pm$ & 0.8 & 0.08 & (-0.21 to 0.21) & 0.983 & 0.829 \\
			22405 & AnteriorThighMuscleL & 5607 & L & [0.7 & , & 3.6] & 1.7 & $\pm$ & 0.5 & 0.06 & (-0.14 to 0.15) & 0.975 & 0.802 \\
			22406 & PosteriorThighMuscleL & 5607 & L & [1.3 & , & 6.3] & 3.4 & $\pm$ & 0.8 & 0.08 & (-0.22 to 0.21) & 0.981 & 0.824 \\
			22407 & VatVolume & 5763 & L & [0.1 & , & 14.4] & 3.8 & $\pm$ & 2.2 & 0.14 & (-0.37 to 0.39) & 0.993 & 0.703 \\
			22408 & AsatVolume & 5763 & L & [1.5 & , & 23.5] & 7.1 & $\pm$ & 3.1 & 0.22 & (-0.62 to 0.61) & 0.990 & 0.822 \\
			22409 & TotalThighMuscle & 5559 & L & [4.3 & , & 19.0] & 10.2 & $\pm$ & 2.5 & 0.18 & (-0.43 to 0.46) & 0.992 & 0.846 \\
			22410 & TotalTrunkFat & 5763 & L & [1.9 & , & 31.6] & 10.9 & $\pm$ & 4.5 & 0.24 & (-0.68 to 0.63) & 0.994 & 0.843 \\
			22415 & TotalAdiposeTissue & 8276 & L & [5.5 & , & 65.9] & 21.1 & $\pm$ & 7.0 & 0.37 & (-0.99 to 1.07) & 0.994 & 0.879 \\
			22416 & TotalLeanTissue & 8276 & L & [12.3 & , & 43.3] & 24.2 & $\pm$ & 4.8 & 0.64 & (-1.92 to 1.69) & 0.963 & 0.846 \\		
			
			&&&&&&&&&&&&&\\
			& \textbf{Other:} & &&&&&&&&&&& \\
			48 & waist & 30441 & cm & [55.0 & , & 184.0] & 88.6 & $\pm$ & 12.6 & 3.35 & (-8.70 to 8.10) & 0.883 & 0.815 \\
			49 & hip & 30443 & cm & [72.0 & , & 157.0] & 101.3 & $\pm$ & 8.6 & 2.60 & (-6.60 to 6.61) & 0.847 & 0.759 \\
			21001 & BMI & 30124 & kg/m$^2$ & [14.2 & , & 62.0] & 26.6 & $\pm$ & 4.3 & 0.41 & (-1.13 to 0.99) & 0.984 & 0.969 \\
			22402 & liverFat & 4419 & \% & [0.0 & , & 46.0] & 4.0 & $\pm$ & 4.7 & 1.35 & (-4.04 to 4.22) & 0.799 & 0.208 \\
			47 & gripStrengthRight & 30053 & kg & [-0.0 & , & 72.0] & 31.3 & $\pm$ & 10.5 & 5.08 & (-12.92 to 12.87) & 0.607 & 0.583 \\
			102 & pulseRate & 25123 & bpm & [33.0 & , & 157.0] & 69.4 & $\pm$ & 12.1 & 8.10 & (-20.13 to 20.76) & 0.262 & 0.058 \\		
			
			\hline
			\hline
			\multicolumn{14}{l}{*SD: Standard deviation, MAE: Mean absolute error, LoA: Limits of agreement.} \\
			\multicolumn{14}{l}{R$^2_{lr}$: Fit of multiple linear regression on age, sex, height and weight.} \\				
		\end{tabular}
	\end{center}
\end{table*}

\begin{table*}[!htb]
	\begin{center}
		\caption{Inference of DXA trunk fields}
		\label{tab_dxa_trunk}		
		\begin{tabular}{
				llrl@{\hskip 0.2cm} 
				r@{\hskip 0.05cm}c@{\hskip 0.05cm}r@{\hskip 0.1cm} 
				r@{\hskip 0.05cm}c@{\hskip 0.05cm}l  | @{\hskip 0.05cm}
				r@{\hskip 0.05cm} c@{\hskip 0.05cm} r|l}
			\hline
			field & name & N & unit &  [min&,&max] & mean &$\pm$& SD & MAE & LoA & R$^2$ & R$^2_{lr}$\\ 
			\hline	
			& \textbf{Android:} & &&&&&&&&&&& \\		
			23244 & BoneMass & 4544 & g & [16.0 & , & 118.0] & 49.0 & $\pm$ & 13.0 & 5.62 & (-14.28 to 14.80) & 0.676 & 0.411 \\
			23245 & FatMass & 4544 & kg & [0.2 & , & 9.4] & 2.5 & $\pm$ & 1.2 & 0.11 & (-0.30 to 0.32) & 0.982 & 0.835 \\
			23246 & LeanMass & 4544 & kg & [0.3 & , & 6.8] & 3.5 & $\pm$ & 0.8 & 0.14 & (-0.34 to 0.35) & 0.945 & 0.833 \\
			23247 & TissueFat & 4544 & \% & [8.1 & , & 65.7] & 39.7 & $\pm$ & 10.5 & 1.29 & (-3.23 to 3.30) & 0.975 & 0.603 \\
			23248 & TotalMass & 4544 & kg & [0.5 & , & 16.4] & 6.0 & $\pm$ & 1.7 & 0.20 & (-0.50 to 0.52) & 0.975 & 0.925 \\
			&&&&&&&&&&&&&\\			
			& \textbf{Gynoid:} & &&&&&&&&&&& \\	
			23261 & BoneMass & 4544 & g & [53.0 & , & 516.0] & 274.1 & $\pm$ & 67.1 & 15.28 & (-37.44 to 40.16) & 0.912 & 0.731 \\
			23262 & FatMass & 4544 & kg & [0.2 & , & 13.5] & 4.2 & $\pm$ & 1.5 & 0.15 & (-0.35 to 0.44) & 0.980 & 0.811 \\
			23263 & LeanMass & 4544 & kg & [0.4 & , & 14.1] & 7.3 & $\pm$ & 1.6 & 0.18 & (-0.49 to 0.45) & 0.977 & 0.876 \\
			23264 & TissueFat & 4544 & \% & [11.8 & , & 61.5] & 36.1 & $\pm$ & 9.1 & 0.92 & (-2.29 to 2.33) & 0.983 & 0.770 \\
			23265 & TotalMass & 4544 & kg & [0.7 & , & 27.5] & 11.8 & $\pm$ & 2.2 & 0.22 & (-0.58 to 0.65) & 0.980 & 0.915 \\
			&&&&&&&&&&&&&\\			
			& \textbf{Trunk:} & &&&&&&&&&&& \\	
			23284 & FatMass & 4544 & kg & [1.2 & , & 46.0] & 14.7 & $\pm$ & 5.9 & 0.46 & (-1.22 to 1.22) & 0.989 & 0.857 \\
			23285 & LeanMass & 4544 & kg & [2.4 & , & 38.6] & 22.8 & $\pm$ & 4.4 & 0.57 & (-1.42 to 1.47) & 0.972 & 0.843 \\
			23286 & TissueFat & 4544 & \% & [10.5 & , & 62.1] & 38.2 & $\pm$ & 9.1 & 0.96 & (-2.24 to 2.53) & 0.982 & 0.647 \\
			23287 & TotalMass & 4544 & kg & [4.2 & , & 82.6] & 38.3 & $\pm$ & 8.6 & 0.71 & (-1.75 to 1.94) & 0.988 & 0.957 \\			
			\hline
			\hline
			\multicolumn{14}{l}{*SD: Standard deviation, MAE: Mean absolute error, LoA: Limits of agreement.} \\
			\multicolumn{14}{l}{R$^2_{lr}$: Fit of multiple linear regression on age, sex, height and weight.} \\				
		\end{tabular}
	\end{center}
\end{table*}

\begin{table*}[!htb]
	\begin{center}
		\caption{Inference of DXA arm fields}
		\label{tab_dxa_arm}		
		\begin{tabular}{
				llrl@{\hskip 0.2cm} 
				r@{\hskip 0.05cm}c@{\hskip 0.05cm}r@{\hskip 0.1cm} 
				r@{\hskip 0.05cm}c@{\hskip 0.05cm}l  | @{\hskip 0.05cm}
				r@{\hskip 0.05cm} c@{\hskip 0.05cm} r|l}
			\hline
			field & name & N & unit &  [min&,&max] & mean &$\pm$& SD & MAE & LoA & R$^2$ & R$^2_{lr}$\\ 
			\hline	
			& \textbf{Left arm:} & &&&&&&&&&&& \\	
			23249 & FatMass & 3834 & kg & [0.4 & , & 4.6] & 1.3 & $\pm$ & 0.5 & 0.12 & (-0.34 to 0.35) & 0.866 & 0.747 \\
			23250 & LeanMass & 3834 & kg & [1.1 & , & 5.5] & 2.6 & $\pm$ & 0.8 & 0.15 & (-0.39 to 0.39) & 0.936 & 0.855 \\
			23251 & TissueFat & 3834 & \% & [11.1 & , & 60.9] & 33.9 & $\pm$ & 9.8 & 1.77 & (-4.45 to 4.39) & 0.947 & 0.770 \\
			23252 & TotalMass & 3834 & kg & [2.0 & , & 8.3] & 4.1 & $\pm$ & 1.0 & 0.25 & (-0.63 to 0.65) & 0.885 & 0.858 \\
			&&&&&&&&&&&&& \\
			& \textbf{Right arm:} & &&&&&&&&&&& \\	
			23253 & FatMass & 3834 & kg & [0.4 & , & 4.6] & 1.4 & $\pm$ & 0.5 & 0.12 & (-0.35 to 0.33) & 0.867 & 0.744 \\
			23254 & LeanMass & 3834 & kg & [1.3 & , & 5.6] & 2.8 & $\pm$ & 0.8 & 0.15 & (-0.38 to 0.40) & 0.940 & 0.860 \\
			23255 & TissueFat & 3834 & \% & [11.0 & , & 60.9] & 33.3 & $\pm$ & 9.6 & 1.68 & (-4.16 to 4.33) & 0.949 & 0.770 \\
			23256 & TotalMass & 3834 & kg & [2.2 & , & 8.3] & 4.3 & $\pm$ & 1.0 & 0.25 & (-0.63 to 0.65) & 0.887 & 0.862 \\
			&&&&&&&&&&&&& \\
			& \textbf{Arms, total:} & &&&&&&&&&&& \\	
			23257 & FatMass & 4544 & kg & [0.4 & , & 9.3] & 2.7 & $\pm$ & 0.9 & 0.23 & (-0.63 to 0.62) & 0.885 & 0.755 \\
			23258 & LeanMass & 4544 & kg & [0.9 & , & 11.0] & 5.4 & $\pm$ & 1.6 & 0.27 & (-0.71 to 0.67) & 0.951 & 0.869 \\
			23259 & TissueFat & 4544 & \% & [11.1 & , & 60.9] & 33.6 & $\pm$ & 9.7 & 1.60 & (-3.87 to 4.12) & 0.955 & 0.775 \\
			23260 & TotalMass & 4544 & kg & [1.3 & , & 16.6] & 8.4 & $\pm$ & 1.9 & 0.44 & (-1.10 to 1.18) & 0.907 & 0.876 \\
			
			\hline
			\hline
			\multicolumn{14}{l}{*SD: Standard deviation, MAE: Mean absolute error, LoA: Limits of agreement.} \\
			\multicolumn{14}{l}{R$^2_{lr}$: Fit of multiple linear regression on age, sex, height and weight.} \\				
		\end{tabular}
	\end{center}
\end{table*}

\begin{table*}[!htb]
	\begin{center}
		\caption{Inference of DXA leg fields}
		\label{tab_dxa_leg}		
		\begin{tabular}{
				llrl@{\hskip 0.2cm} 
				r@{\hskip 0.05cm}c@{\hskip 0.05cm}r@{\hskip 0.1cm} 
				r@{\hskip 0.05cm}c@{\hskip 0.05cm}l  | @{\hskip 0.05cm}
				r@{\hskip 0.05cm} c@{\hskip 0.05cm} r|l}
			\hline
			field & name & N & unit &  [min&,&max] & mean &$\pm$& SD & MAE & LoA & R$^2$ & R$^2_{lr}$\\ 
			\hline					
			& \textbf{Left leg:} & &&&&&&&&&&& \\
			23266 & FatMass & 3834 & kg & [0.9 & , & 13.6] & 3.9 & $\pm$ & 1.5 & 0.18 & (-0.52 to 0.44) & 0.972 & 0.746 \\
			23267 & LeanMass & 3834 & kg & [3.7 & , & 16.1] & 7.9 & $\pm$ & 1.8 & 0.29 & (-0.68 to 0.81) & 0.956 & 0.879 \\
			23268 & TissueFat & 3834 & \% & [10.7 & , & 62.0] & 32.5 & $\pm$ & 9.5 & 1.06 & (-2.61 to 2.68) & 0.980 & 0.767 \\
			23269 & TotalMass & 3834 & kg & [7.0 & , & 25.6] & 12.3 & $\pm$ & 2.4 & 0.38 & (-0.97 to 1.07) & 0.952 & 0.871 \\
			&&&&&&&&&&&&&\\
			& \textbf{Right leg:} & &&&&&&&&&&& \\
			23270 & FatMass & 3834 & kg & [1.0 & , & 13.5] & 3.9 & $\pm$ & 1.5 & 0.17 & (-0.48 to 0.47) & 0.974 & 0.742 \\
			23271 & LeanMass & 3834 & kg & [4.0 & , & 15.4] & 8.1 & $\pm$ & 1.9 & 0.28 & (-0.71 to 0.74) & 0.960 & 0.883 \\
			23272 & TissueFat & 3834 & \% & [10.7 & , & 61.9] & 32.6 & $\pm$ & 9.5 & 1.02 & (-2.57 to 2.56) & 0.981 & 0.763 \\
			23273 & TotalMass & 3834 & kg & [6.8 & , & 25.6] & 12.5 & $\pm$ & 2.4 & 0.36 & (-0.92 to 1.01) & 0.957 & 0.875 \\
			&&&&&&&&&&&&&\\
			& \textbf{Legs, total:} & &&&&&&&&&&& \\
			23274 & FatMass & 4544 & kg & [1.0 & , & 27.1] & 7.8 & $\pm$ & 3.0 & 0.30 & (-0.79 to 0.89) & 0.979 & 0.750 \\
			23275 & LeanMass & 4544 & kg & [3.4 & , & 31.5] & 16.0 & $\pm$ & 3.7 & 0.48 & (-1.14 to 1.33) & 0.970 & 0.888 \\
			23276 & TissueFat & 4544 & \% & [10.7 & , & 62.0] & 32.6 & $\pm$ & 9.5 & 0.92 & (-2.19 to 2.42) & 0.985 & 0.768 \\
			23277 & TotalMass & 4544 & kg & [4.5 & , & 52.7] & 24.8 & $\pm$ & 4.8 & 0.63 & (-1.81 to 1.53) & 0.967 & 0.880 \\			
			\hline
			\hline
			\multicolumn{14}{l}{*SD: Standard deviation, MAE: Mean absolute error, LoA: Limits of agreement.} \\
			\multicolumn{14}{l}{R$^2_{lr}$: Fit of multiple linear regression on age, sex, height and weight.} \\				
		\end{tabular}
	\end{center}
\end{table*}

\begin{table*}[]
	\begin{center}
		\caption{Symmetrical measurements}
		\label{tab_symmetry}		
		\begin{tabular}{llll|l|rrr}
			\hline
			field$_a$ & name$_a$ & field$_b$ & name$_b$ & N & $r_{(a,b)}$ & $r_{(a, net)}$ & $r_{(b, net)}$ \\ 
			\hline
			22405 & MriAnteriorThighLeanMuscleLeft & 22403 & MriAnteriorThighLeanMuscleRight & 5559 & 0.974 & \textbf{0.988} & \textbf{0.987} \\
			22406 & MriPosteriorThighLeanMuscleLeft & 22404 & MriPosteriorThighLeanMuscleRight & 5559 & 0.983 & \textbf{0.991} & \textbf{0.992} \\
			&&&&&&&\\
			23249 & DxaArmFatMassLeft & 23253 & DxaArmFatMassRight & 3834 & \textbf{0.971} & 0.931 & 0.931 \\
			23250 & DxaArmLeanMassLeft & 23254 & DxaArmLeanMassRight & 3834 & \textbf{0.978} & 0.968 & 0.969 \\
			23251 & DxaArmTissueFatPercentageLeft & 23255 & DxaArmTissueFatPercentageRight & 3834 & \textbf{0.984} & 0.973 & 0.974 \\
			23252 & DxaArmTotalMassLeft & 23256 & DxaArmTotalMassRight & 3834 & \textbf{0.966} & 0.941 & 0.942 \\		
			&&&&&&&\\
			23266 & DxaLegFatMassLeft & 23270 & DxaLegFatMassRight & 3834 & \textbf{0.989} & 0.986 & 0.987 \\
			23267 & DxaLegLeanMassLeft & 23271 & DxaLegLeanMassRight & 3834 & \textbf{0.984} & 0.978 & 0.980 \\
			23268 & DxaLegTissueFatPercentageLeft & 23272 & DxaLegTissueFatPercentageRight & 3834 & \textbf{0.992} & 0.990 & 0.990 \\
			23269 & DxaLegTotalMassLeft & 23273 & DxaLegTotalMassRight & 3834 & \textbf{0.982} & 0.976 & 0.979 \\
			\hline
			\hline
			\multicolumn{8}{l}{*Correlations between symmetrical fields and network predictions. The fields (a) and (b) correlate by $r_{(a,b)}$  whereas the network}\\
			\multicolumn{8}{l}{  output correlates to field (a) by $r_{(a, net)}$. Only those N subjects were evaluated for whom both measurements were available.} \\	
			\multicolumn{8}{l}{Bold font denotes numerically higher values.}		
		\end{tabular}
	\end{center}
\end{table*}

\end{document}